# Classical dynamical localization in a strongly driven two-mode mechanical system

Hao Fu,[1] Zhi-cheng Gong,[1] Li-ping Yang,[2] Tian-hua Mao,[1] Chang-pu Sun,[2,3,4] Su Yi,[4,5] Yong Li,[2,3,4*] & Geng-yu Cao[1,*]

[1] State Key Laboratory of Magnetic Resonance and Atomic and Molecular Physics, Wuhan Institute of Physics and Mathematics, Chinese Academy of Sciences, Wuhan 430071, P. R. China

[2] Beijing Computational Science Research Center, Beijing 100193, China

[3] Synergetic Innovation Center of Quantum Information and Quantum Physics, University of Science and Technology of China, Hefei, Anhui 230026, China

[4] Synergetic Innovation Center for Quantum Effects and Applications, Hunan Normal University, Changsha 410081, China

[5] Institute of Theoretical Physics, Chinese Academy of Sciences, Beijing 100190, China

*Correspondence and requests for materials should be addressed to Y. L. (liyong@csrc.ac.cn) or G.-Y. C. (gycao@wipm.ac.cn)

**Abstract**

We report the realization of dynamical localization in a strongly driven two-mode optomechanical system consisting of two coupled cantilevers. Due to the coupling, mechanical oscillations can transport between the cantilevers. However, by placing one of the cantilevers inside a harmonically oscillating optical trap, we demonstrate that mechanical oscillations become tightly bounded to the isolated cantilevers rather than propagate away at specific driving parameters. The effect of dynamical localization is employed to induce a conductor-to-insulator-like transition in the two-mode mechanical system, which opens up new possibilities for further coherent control of transport phenomena in coupled-mechanical-resonator based lattice.

**Introduction**

Coherent manipulation of quantum states through a periodic driving field has been extensively studied in various systems. In connection with non-adiabatic crossing of energy levels, the non-trivial phenomena in periodic transition of avoided-crossing point have been investigated in terms of Landau-Zener-Stückelberg (LZS) interferometry, through which the coupling strength between two basis states can be parametrically tuned [1]. Particularly, when the coupling vanishes, the distinctive cases of dynamical localization (DL) and coherent destructive tunneling (CDT) have attracted long-lasting interests for their importance in coherent control of transport phenomena [2-4]. For example, the DL, as initially studied for charged particle in a harmonically oscillating electric field [4], has been recently achieved in curved photonic lattices for coherent control of light propagation [5-8]. And the striking effect of CDT, which can be regarded as the DL in fast driving limit, has been explored for applications such as photon shuttle, coherent switcher and coupler [9-13].

In the past few years, nano-mechanical resonators with extremely small mass and unprecedented low dissipation have been fabricated for ultra-precise sensing and even room-temperature quantum mechanics experiments [14-17]. The cutting-edge technology opens up new possibilities of realizing mechanical lattice with coupled mechanical resonators for quantum information processing and high-sensitivity signal transducers [18-23]. The broad applications of multi-mode mechanical system intrigue intensive interests on coherent manipulation of mechanical oscillations. For example, two-mode mechanical system has been demonstrated as a classical analogy of quantum two-level system, where the Rabi physics was applied for full coherent control of mechanical oscillations in Bloch sphere [24-26]. And more recently, studies in the two-mode mechanical system have shown that non-adiabatic transition of the avoided-crossing point follows the well-known Landau-Zener (LZ) model, in which coherent splitting and recombination of mechanical oscillations have been achieved through single and round-trip LZ transition respectively [27-29]. With the improved understanding on the non-adiabatic dynamics of the two-mode mechanical system, it now becomes more feasible to realize a classical LZS interferometry and in particular a mechanical analogy of DL for further coherent control of transport phenomena in the mechanical lattice.

In this letter, we present the DL in a strongly driven two-mode mechanical system consisting of two coupled fundamental flexural cantilever modes. By trapping one of the cantilevers inside a fiber-based Fabry-Pérot cavity and modulating the depth of the optical trap, the avoided-crossing point can be traversed periodically for a LZS-like interferometry. In order to realize the DL, the dynamics of tunneling oscillation between the cantilevers subjected to a harmonically oscillating driving field is investigated at different driving parameters. We demonstrate that the DL can be realized at specific driving parameters as a result of destructive interference of mechanical oscillations. The effect of DL is employed to induce a conductor-to-insulator-like transition in the two-mode mechanical system and explored further for its perspective applications on coherent control of mechanical lattice based devices.

**Experiments**

The two cantilevers used in our experiment are elastically coupled by connecting to the same overhang. Due to a slight difference in their geometries, the fundamental flexural motions $x_j$ ($j$=1, 2) of the two cantilevers are non-degenerated with their bare frequencies $\omega_1/2\pi = 6,817.5$ Hz and $\omega_2/2\pi = 7,203.5$ Hz respectively (see supplementary materials). In high vacuum condition, the energy dissipation rates of the cantilevers are approximately the same $\gamma_j/2\pi = 5.2$ Hz. As schematically illustrated in Fig. 1(a), we insert cantilever 1 into a low finesse fiber-based Fabry-Pérot cavity to form a membrane-in-the-middle optomechanical system, in which the cantilever is trapped by a 1064 nm beam through dispersive coupling between the mechanical mode and the optical mode [30,31]. In the presence of the optical trap, the effective frequency of cantilever 1 becomes trapping power $P$ dependent $\omega_{eff,1}^2 = \omega_1^2 + gP$ with $g \approx 1.54 \times 10^6\ rad^2 \cdot s^{-2} \cdot \mu W^{-1}$ denoting the optical trapping strength; while the frequency of cantilever 2 remains unaffected.

Therefore, hybridization between the two cantilevers can be mediated optically by tuning the mode mismatch. Due to the elastic coupling between the two cantilevers, the motions of the cantilevers are hybridized into two normal modes with the two cantilevers oscillating in-phase and out-of-phase respectively [27]. In our experiment, the motion of cantilever 1 is monitored by a weak 1310 nm probe. As shown in Fig.

1(b), the two normal modes can be distinguished in frequency domain. For the large intrinsic mode mismatch ($|\omega_{eff,1} - \omega_2| \gg \Delta$) at $P = 0$, the cantilevers are weakly hybridized and the normal modes are localized on the corresponding cantilevers. When the trapping power is adiabatically ramped to $P_{ac} = 139$ µW, the deep mode hybridization at the degenerated point gives rise to two delocalized normal modes and an avoided crossing of the modes is clearly observed. The coupling strength between the cantilevers measured as anti-crossing gap $\Delta/2\pi = 113$ Hz indicates that the system works in strong coupling regime ($\Delta \gg \gamma_j$).

**Results and discussions**

The strong coupling allows coherent tunneling of mechanical oscillations between the isolated cantilevers, which can be manifested as an intensive Rabi-like oscillation at the avoided-crossing point [26,27]. When a harmonically oscillating optical trap is applied to modulate the effective frequency of cantilever 1 around the degenerated point $\omega_{eff,1}^2(t) = \omega_2^2 + \alpha_d^2 \sin(\omega_d t)$, the motion of the cantilevers can be written approximately as $x_j(t) \approx Re[a_j(t)e^{-i\omega_2 t}] = |a_j(t)|\sin[\omega_2 t + \phi_j(t)]$ by neglecting all off-resonance idler oscillations, where $|a_j(t)|$ and $\phi_j(t)$ represent the slowly-varying amplitude and oscillation phase respectively (see supplementary materials). Therefore, the tunneling dynamics of the cantilevers can be described by

$$\begin{bmatrix} i\frac{d}{dt} + \varepsilon(t) + i\frac{\gamma_1}{2} & \frac{\Delta}{2} \\ \frac{\Delta}{2} & i\frac{d}{dt} + i\frac{\gamma_2}{2} \end{bmatrix} \begin{bmatrix} a_1(t) \\ a_2(t) \end{bmatrix} = 0.$$

Here, in similar with periodically driven double-well systems [4], $\varepsilon(t) = A_d \sin(\omega_d t)$ is referred as the driving field with the driving amplitude $A_d = \frac{\alpha_d^2}{2\omega_2}$. As a result of the periodic driving, the coupling strength between the cantilevers becomes parametrically tunable with the effective coupling rate $\Delta_{eff} = \Delta \exp\left[\pm i \frac{A_d}{\omega_d} \cos(\omega_d t)\right]$. In high-frequency regime ($\omega_d \gg \Delta$), the effective coupling rate can be calculated as $\Delta_{eff} = \Delta |J_0(A_d/\omega_d)|$ by neglecting all fast oscillating terms, where $J_0(z)$ is the zero-order Bessel function of the first kind [32-34].

In our experiments, the trapping power is modulated $P(t) = P_{ac}[1 - \sin(\omega_d t)]$ to drive the system around the avoided-crossing point with a fixed amplitude of $A_d/2\pi = 375$ Hz. As showed in Fig. 2(a), the motion of the coupled cantilevers in the harmonically oscillating driving field has two Floquet modes. Here, the effective coupling strength measured as the anti-crossing gap is parametrically tuned by changing the driving frequency $\omega_d$. For $\omega_d/2\pi = 1,000$ Hz, the coupling between the cantilevers is nearly turned on completely with $\Delta_{eff} = 2\pi \times 111$ Hz $\approx \Delta$ [Fig. 2(b)]. And the equal height of resonances indicates that the two Floquet modes contribute identically to the motion of each cantilever. With decreasing $\omega_d$, the two Floquet modes move close to each other until the driving frequency reaches a critical value $\omega_d/2\pi = 167$ Hz, where the coupling between the cantilevers is turned off as the anti-crossing gap vanishes; further decreasing $\omega_d$ opens the anti-crossing gap again. Experimental results show that the anti-crossing gap exhibits a Bessel-function-like dependence on the driving parameter $A_d/\omega_d$ in the high-frequency regime. With the driving frequency turned down, however, we notice that the relation starts to deviate from the Bessel function. The observed difference can be attributed to the higher-harmonic oscillating terms, which cannot be simply neglected in the intermediate-frequency regime ($\omega_d \sim \Delta$).

In addition to the realization of full parametric control over the effective coupling rate, we demonstrate that the tunneling dynamics in the strongly driven two-mode system can be mediated consequently. For this purpose, the coupled cantilevers, which are thermally occupied at room temperature, is initialized by piezo-electrically actuating the lower-frequency in-phase normal mode at the trapping power of $P_0 = 350\ \mu$W, where almost all oscillation energy is localized on cantilever 2 due to a large mode mismatch. Immediately after the initial state preparation, the system is driven by the control pulse as schematically illustrated in Fig. 3(a). First of all, we diabatically ramp the trapping power to $P_{ac}$ so that most mechanical oscillations remain on cantilever 2 ($|a_2(0)| \gg |a_1(0)|$). Then, the driving field as described above is applied for time $t_d$. For mechanical oscillations undergoing multi-passage LZ transition, repeated coherent splitting and recombination of mechanical oscillations leads to a tunneling oscillation, a LZS-like interference fringe in time domain. The final amplitude of the cantilevers $|a_j(t_d)|$ is readout after diabatically ramping the trapping power back to $P_0$ again. The rescaled amplitude of cantilever 1 $\frac{|a_1(t_d)|}{|a_2(0)|}$ is plotted in Fig. 3(b). For $\omega_d/2\pi = 1,000$ Hz, the strong tunneling oscillation at the frequency of 109 Hz confirms that the coupling between the cantilevers is turned on. As the driving frequency decreasing to 200 Hz, the tunneling oscillation slows down gradually. And, as expected, the frequency of tunneling oscillation agrees well with the anti-crossing gap measured in Fig. 2(a).

Specifically, the tunneling oscillation will be infinitely slow when the coupling is turned off at specific driving parameters. Indeed, as the anti-crossing gap vanishes at $\omega_d/2\pi = 167$ Hz, the tunneling oscillation in Fig. 3(c) becomes unobservable and the oscillation energy on cantilever 2 exhibits an exponential decay. Nevertheless, numerical results of the mechanical oscillations distribution probability, which is calculated as oscillation power $\left(\frac{|a_j(t_d)|}{|a_2(0)|}\right)^2$, shows that the mechanical oscillations are not frozen on cantilever 2 over all the driving intervals. Instead, a fraction of the mechanical oscillations can still tunnel to cantilever 1 when the avoided-crossing point is traversed [27,28]. However, as a consequence of destructive interference at the following LZ transition, the mechanical oscillations return back to cantilever 2 again after each full driving cycle and hence the weak fluctuation on the oscillation power at the driving frequency in Fig. 3(c). The phenomenon that mechanical oscillations are tightly bounded to the initialized cantilever rather than propagate away can be regarded as the realization of the DL [4]. And we remark that the tunneling probability during each driving cycle can be significantly suppressed when the avoided-crossing point is traversed diabatically. The enhanced localization is confirmed by further numerical calculation using a stronger driving field ($A_d/2\pi = 1,000$ Hz). In limit of strong and high-frequency driving, oscillation can be frozen on the cantilever where it is initially localized as a result of CDT at $\Delta_{eff} = 0$ [2,9].

Here, we emphasize that although large mode mismatch can lead to localized oscillation as well [35], the DL originates from the superposition of degenerated Floquet modes. The advantageous nature of DL ensures some sophisticated manipulations of mechanical oscillations such as coherent switching the tunneling oscillation. As a proof-of-the-principle demonstration, the tunneling oscillation is switched ON-OFF-ON by a composite driving pulse as schematically illustrated in Fig. 4(a). The pulse segments with the driving frequency of 1,000 Hz and 167 Hz are used to switch on and off the tunneling oscillation respectively. After initializing the mechanical oscillations to cantilever 2, we apply the switch-on segment to switch on the tunneling oscillation first. As shown in Fig. 4(b), the tunneling oscillation can be paused at an arbitrary phase by the following switch-off segment until the second switch-on segment arrives. The resumed strong tunneling oscillation with a consistent phase after the switch-off segment is a clear evidence of well-preserved phase coherence, which can be recognized as the maintaining of the superposition of the Floquet modes during the switching operation.

**Conclusions**

In conclusion, we have demonstrated the DL in a strongly driven two-mode system consisting of two elastically coupled micro-cantilevers. Complete control of the effective coupling rate between the cantilevers has been achieved as result of LZS-like interference of mechanical oscillations. We showed that mechanical oscillations can be dynamically localized to isolated cantilevers by turning off their effective coupling. Moreover, as one of the applications of the DL, switch of the tunneling oscillation with well-preserved phase coherence has been achieved experimentally by inducing a conductor-to-insulator-like transition at the DL condition. As a hallmark of DL, our further analysis on the coupled-ten-resonator based mechanical lattice revealed that the distribution of oscillation power on the lattice manifests an archetypal exponential decay away from the resonator where it is initially localized (see supplementary materials). Therefore, in applications where the tunneling probability should be kept remaining small during the operation, we propose that technically demanding requirement on CDT can be greatly relaxed by simply adding only a few additional resonators, which makes high performance mechanical devices in a more parametrically feasible regime. Prospectively, the realization of the DL might offer a promising approach for high degree coherent control of transport phenomena in coupled-mechanical-resonator based lattices for broad applications such as coherent switcher, signal splitter, tunable coupler, and non-reciprocal transducer.

**Acknowledgements**

This work was supported by National Natural Science Foundation of China (Grants No. 91636220, No. 11421063, and No. U1530401), China Postdoctoral Science Foundation (Grant No. 2015M580966), and National Key Research and Development Program of China (Grant No. 2016YFA0301200). We would like to thank nanofabrication facility in Suzhou Institute of Nanotech and Nano bionics (CAS) for fabricating the micro-cantilevers.

**Figures and captions**

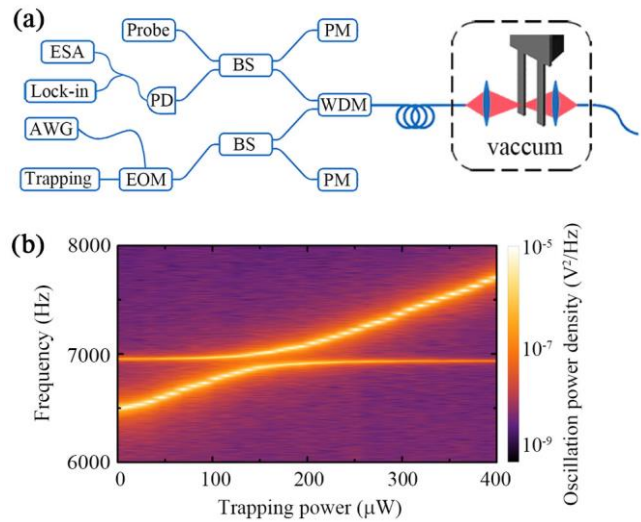

Figure 1 (a) Schematics of experimental setup. Control pulse from an arbitrary waveform generator (AWG) is used to modulate the intensity of the 1064 nm control beam through an electro-optical modulator (EOM). The power of control beam is monitored after a beam splitter (BS) by a power meter (PM). An additional weak 1310 nm probe of 2 μW is employed to monitor the motion of the cantilever inside the fiber-based cavity. The intensity of reflected probe, which is divided from the control beam via a wavelength-division multiplexing (WDM), is recorded by a photodetector (PD) and then analyzed in time and frequency domain using lock-in amplifier (LIA) and electro-spectral analyzer (ESA) respectively. (b) Thermal oscillation power spectral density of the coupled cantilevers in static optical trap. The upper and lower normal modes correspond to the out-of-phase and in-phase motion of cantilevers.

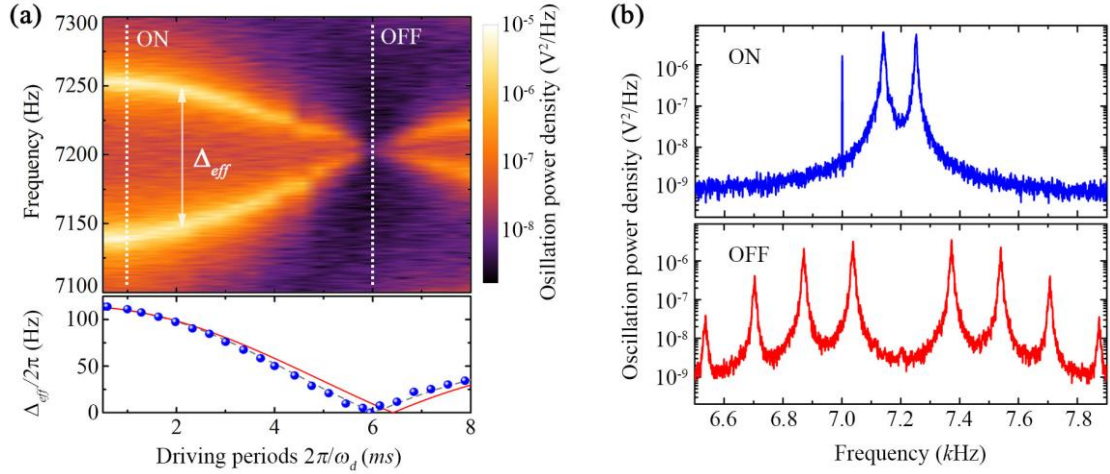

Figure 2 (a) Thermal oscillation power spectrum of coupled cantilevers subjected to harmonic oscillating optical trap. The driving parameters for turning on and off the coupling are marked by dotted lines in upper panel. The measured anti-crossing gap (dots) is plotted in lower panel with the effective coupling rate calculated from $\Delta_{eff} = \Delta |J_0(A_d/\omega_d)|$ (solid line). The observed deviation in intermediate-frequency regime is confirmed by numerical results (dashed line) using the same parameters in the experiments. (b) Thermal oscillation power spectral density of the coupled cantilevers when the coupling is turned on (upper) and off (lower) as marked in (a). The sharp peak at 7.0 $k$Hz in the upper panel originates from the higher-order harmonics of driving field.

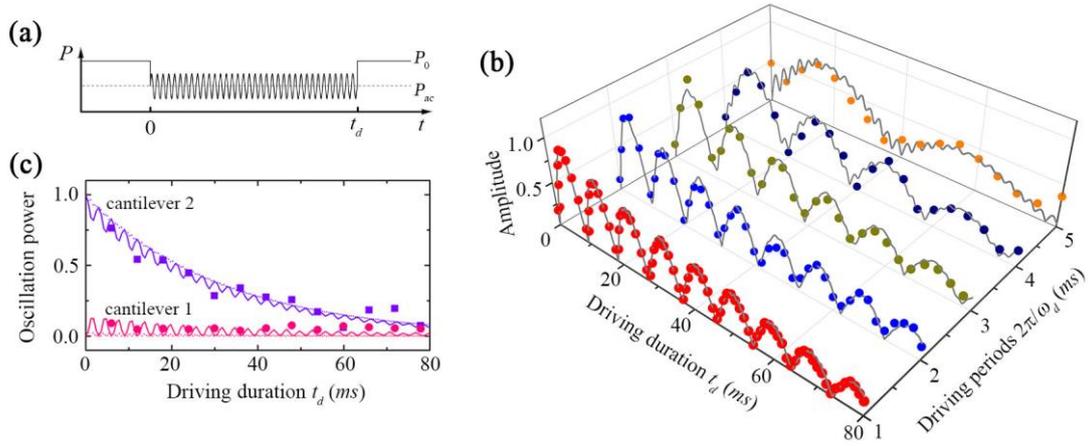

Figure 3 (a) Scheme of control pulse for observation of the LZS interference in time domain. The driving field with $A_d/2\pi = 375$ Hz is applied for an integral numbers of driving cycles. (b) Rescaled oscillation amplitude of cantilever 1 for different driving periods. The experimental measurements after an integral numbers of driving cycles (dots) are plotted with numerical results (solid lines). (c) Rescaled oscillation power of cantilevers at DL condition. When a driving field of $\omega_d/2\pi = 167$ Hz is applied to turn off the coupling, the measured data of both cantilever 1 (dots) and 2 (rectangle) are plotted with numerical results (solid lines). For comparison, the DL condition under the driving field of $A_d/2\pi = 1,000$ Hz is numerically calculated as well (dashed lines).

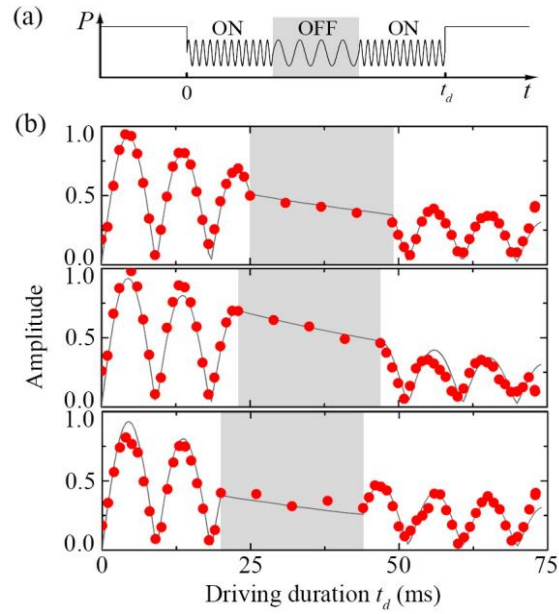

Figure 4 (a) Scheme of control pulse for switcher-like control of tunneling oscillation. The system is initialized to cantilever 2. The switch-off segment, which is shadowy marked, is sandwiched between two switch-on segments. All of the three pulse segments have the same driving amplitude $A_d/2\pi = 375$ Hz and consist of integral numbers of driving cycles. (b) Rescaled oscillation amplitude of cantilever 1 versus the driving duration $t_d$. Experimental measurements (dots) are plotted with numerical results (solid lines), in which the weak tunneling oscillation at the driving frequency $\omega_d$ is filtered out for clarity. The switch-off region is shadowy marked accordingly.

Supplementary Material

I. **Effective coupling of the strongly driven coupled two cantilevers**

The single-crystal silicon cantilevers used in our experiments are fabricated on silicon-on-insulator wafer with the dimension of 210 μm × 10 μm × 220 nm (cantilever 1) and 205 μm × 10 μm × 220 nm (cantilever 2) respectively. Due to the small difference in their geometry, the fundamental flexural modes of the cantilevers are non-degenerated. For simplicity and clarity, we assume their effective masses are identical ($m_1 \approx m_2 = m$) and attribute the intrinsic mode mismatch fully to the difference in their spring constants ($k_1 \neq k_2$). By inserting cantilever 1 inside a fiber-based Fabry-Pérot cavity to form a membrane-in-the-middle optomechanical system and tuning the cavity length so that dispersive optomechanical coupling is dominant, the cantilever is optically trapped through introducing an additional optical spring $k_{opt} = mgP$ with $g$ representing the dispersive optomechanical coupling strength [S1-S3]. As a result, the effective frequency of cantilever 1 becomes trapping power $P$ dependent $\omega_{eff,1}^2 = \frac{k_1 + k_{opt}}{m} = \omega_1^2 + gP$; while the frequency of cantilever 2, $\omega_2^2 = \frac{k_2}{m}$, remains unaffected. The two cantilevers separated by 20 μm are elastically coupled by connecting to the same thin silicon overhang extended 10 μm out of the insulator substrate. Therefore, the motions of the two cantilevers $x_j(t)$ ($j = 1,2$) can be described as

$$\begin{pmatrix} \frac{d^2}{dt^2} + \gamma_1 \frac{d}{dt} + \omega_{eff,1}^2 & J \\ J & \frac{d^2}{dt^2} + \gamma_2 \frac{d}{dt} + \omega_2^2 \end{pmatrix} \begin{pmatrix} x_1 \\ x_2 \end{pmatrix} = 0, \quad (S1)$$

where $\gamma_j$ and $J$ denote the damping rate and coupling strength respectively. In the case of strong coupling, we can diagonalize Eq. S1 by neglecting the effects of the damping. As a result, we obtain the out-of-phase ($X_+$) and in-phase ($X_-$) motions normal modes $\begin{pmatrix} X_+ \\ X_- \end{pmatrix} = U \begin{pmatrix} x_1 \\ x_2 \end{pmatrix}$ with $U = \begin{pmatrix} \cos\frac{\theta}{2} & \sin\frac{\theta}{2} \\ -\sin\frac{\theta}{2} & \cos\frac{\theta}{2} \end{pmatrix}$, where $\theta$ satisfies $\tan\theta \approx \frac{\Delta}{\omega_{eff,1} - \omega_2}$. Therefore, the motion of each cantilever is a superposition of two normal modes. And the eigen-frequencies of the normal modes can be calculated

$$\omega_\pm^2 = \frac{1}{2}(\omega_{eff,1}^2 + \omega_2^2 \pm \sqrt{(\omega_{eff,1}^2 - \omega_2^2)^2 + 4J^2}). \tag{S2}$$

At the point where the mode mismatch vanishes ($\omega_{eff,1} = \omega_2$), avoided crossing of mechanical resonances can be observed in Fig. 1(b) and the elastic coupling strength can be measured as the anti-crossing gap $\Delta = \omega_+ - \omega_- \approx J/\omega_2$ when $\Delta \ll \omega_2$. The measured anti-crossing gap $\Delta = 2\pi \times 113$ Hz $\gg \gamma_j$ confirms that the system is indeed in strong coupling regime. By fitting the resonant frequencies of two normal modes to Eq. S2, the bare frequencies of two cantilevers are obtained with $\omega_1/2\pi = 6{,}817.5$ Hz and $\omega_2/2\pi = 7{,}203.5$ Hz.

When a time-varying trapping power $P(t) = P_0 + P_d \sin(\omega_d t)$ is applied to modulated the depth of the optical trap, the effective frequency of cantilever 1 becomes time-dependent $\omega_{eff,1}^2(t) = \omega_0^2 + \alpha_d^2 \sin(\omega_d t)$ with $\omega_0^2 = \omega_1^2 + gP_0$ and $\alpha_d^2 = gP_d$. Generally, the motions of the cantilevers in the harmonically-oscillating optical trap can be written by taking the $n$-th ($n = 0, \pm 1, \ldots$) idler amplitudes $a_1^{(n)}$ of cantilever 1 into account as

$$x_1(t) = Re\left[\sum_n a_1^{(n)}(t) e^{-i(\omega_0 + n\omega_d)t}\right]$$
$$x_2(t) = Re[a_2(t) e^{-i\omega_2 t}], \tag{S3}$$

where $a_2(t)$ and $a_1^{(n)}(t)$ represent the oscillation amplitudes [S4]. Then, the motion

equation becomes

$$\begin{pmatrix} \frac{d^2}{dt^2} + \gamma_1 \frac{d}{dt} + \omega_{eff,1}^2(t) & J \\ J & \frac{d^2}{dt^2} + \gamma_2 \frac{d}{dt} + \omega_2^2 \end{pmatrix} \begin{pmatrix} \sum_n a_1^{(n)}(t) e^{-i(\omega_0 + n\omega_d)t} \\ a_2(t) e^{-i\omega_2 t} \end{pmatrix} = 0 \ .$$

(S4)

Specifically, when the intrinsic frequency of cantilever 1 is modulated around the degenerated point ($\omega_0 = \omega_2$), $\omega_{eff,1}^2(t) = \omega_2^2 + \alpha_d^2 \sin(\omega_d t)$, we can neglect all the off-resonance terms ($n \neq 0$) in Eq. S4 providing $\omega_d > \Delta_{eff}$, which is the case for the full parametrical range in our experiments [S5]. Therefore, the tunneling oscillation can be treated as a zero-order process, a resonant effect with $n = 0$. In this case, although the motion of cantilever 1 in Eq. S3 includes higher-order idler oscillations, its motion is dominant by the term with $n = 0$ because the mechanical oscillations on cantilever 2 cannot be transferred to those off-resonance higher-order components, which remain as thermally occupied at room temperature. Thus, the motions of the cantilevers can be expressed approximately as

$$x_1(t) \approx Re[a_1(t) e^{-i\omega_2 t}] = |a_1(t)| \sin[\omega_2 t + \phi_1(t)]$$

$$x_2(t) = Re[a_2(t) e^{-i\omega_2 t}] = |a_2(t)| \sin[\omega_2 t + \phi_2(t)], \quad \text{(S5)}$$

where $|a_j(t)|$ is the slow varying oscillating amplitude and the oscillation phase satisfies $\tan\phi(t) = \frac{Re[a_j(t)]}{Im[a_j(t)]}$ with $Re[a_j(t)]$ and $Re[a_j(t)]$ denoting the real and imaginary part of $a_j(t)$. Since the frequency of tunneling oscillation is small as compared to the resonant frequency of cantilever ($\omega_2 \gg \Delta_{eff}$), Eq. S4 can be reduced to

$$\begin{bmatrix} (\gamma_1 - 2i\omega_2)\frac{d}{dt} + \alpha_d^2 \sin(\omega_d t) - i\gamma_1 \omega_2 & J \\ J & (\gamma_1 - 2i\omega_2)\frac{d}{dt} - i\gamma_2 \omega_2 \end{bmatrix} \begin{bmatrix} a_1(t) \\ a_2(t) \end{bmatrix} = 0, \quad \text{(S6)}$$

In the case of $\omega_2 \gg \gamma_j$, Eq. S6 can be further reduced to

$$\begin{bmatrix} -i\frac{d}{dt} + \varepsilon(t) - i\frac{\gamma_1}{2} & \frac{\Delta}{2} \\ \frac{\Delta}{2} & -i\frac{d}{dt} - i\frac{\gamma_1}{2} \end{bmatrix} \begin{bmatrix} a_1(t) \\ a_2(t) \end{bmatrix} = 0, \quad (S7)$$

where $\varepsilon(t) = A_d \sin(\omega_d t)$ with amplitude $A_d = \frac{\alpha_d^2}{2\omega_2}$ representing the driving field. Performing the transformation $a_1(t) = b_1(t)e^{-i\int \varepsilon(t)dt}$ and $a_2(t) = b_2(t)$, it yields

$$\begin{bmatrix} -i\frac{d}{dt} - i\frac{\gamma_1}{2} & \frac{\Delta}{2}e^{i\int A_d \sin(\omega_d t)dt} \\ \frac{\Delta}{2}e^{-i\int A_d \sin(\omega_d t)dt} & -i\frac{d}{dt} - i\frac{\gamma_1}{2} \end{bmatrix} \begin{bmatrix} b_1(t) \\ b_2(t) \end{bmatrix} = 0, \quad (S8)$$

which indicates that the effective coupling is rescaled by $\exp\left[\pm i\frac{A_d}{\omega_d}\cos(\omega_d t)\right] = \sum_n (\pm 1)^n J_n(\frac{A_d}{\omega_d}) \exp[\pm in\omega_d t]$. In high-frequency regime ($\omega_d \gg \Delta$), we have $\Delta_{eff} = \Delta \left| J_0(\frac{A_d}{\omega_d}) \right|$ by neglecting all fast oscillating terms ($n \neq 0$) [S6-S8].

In this letter, the time-domain LZS interference fringe is calculated by numerically solving Eq. S7 with the initial state $\begin{bmatrix} a_1(0) \\ a_2(0) \end{bmatrix} = \begin{bmatrix} 0 \\ 1 \end{bmatrix}$. And the numerical results using the exact parameters in the experiments are plotted with experimental data in the main text. As a comparison, switching off the effective coupling using a driving field ($A_d/2\pi = 1,000$ Hz) is also analyzed. Numerical result shows that the effective coupling between cantilevers is switched off when $\omega_d/2\pi = 419.3$ Hz. And, as expected, the tunneling probability during each driving cycle is highly suppressed. The phenomenon that mechanical oscillations are frozen on the initial site can be treated as a consequence of coherent destructive tunneling (CDT) at the strong driving condition.

## II. Dynamical localization in coupled-ten-resonator based mechanical lattice

We further consider a coupled-ten-resonator based mechanical lattice as illustrated in Fig. S1(a). Each resonator is coupled to the nearest ones only. The motion of the $j$-th ($j = 1, \ldots, 10$) resonator can be described as

$$-i\dot{a}_j(t) + \varepsilon_j(t)a_j(t) + \frac{\Delta}{2}\left[a_{j+1}(t) + a_{j-1}(t)\right] = 0 \tag{S9}$$

where $\Delta$ represents the coupling strength to the nearest resonators. We selectively drive resonator 1, $\varepsilon_1(t) = A_d \sin(\omega_d t)$ and $\varepsilon_j(t) = 0$ ($j = 2, \ldots, 10$). For the initial state $\{a_1(0), a_2(0), \ldots, a_{10}(0)\} = \{1, 0, \ldots, 0\}$, time-averaged oscillation power $\langle|a_j|\rangle^2$ at the DL condition is numerically calculated the from Eq. S9. As shown in Fig. S1(b), the energy distribution on the mechanical lattice exhibits an exponential decay with the resonator number $\langle|a_j|\rangle^2 \propto e^{-\frac{j-1}{L}}$, where $L$ denotes the localization length. When we increase the driving strength $A_d/2\pi$ from 375 Hz to 1,000 Hz, the localized effect is enhanced significantly with $L$ decreasing from 0.42 to 0.22. Taking a mechanical lattice consisting of two coupled resonators for example, it indicates that the tunneling probability can be suppressed by nearly one order when the driving amplitude increases from 375 Hz to 1,000 Hz. However, we point out that a similar result can be also realized at $A_d/2\pi = 375$ Hz by, for example, adding only one additional resonator for a coupled-three-resonator based mechanical lattice, which relaxes the demanding requirement on strong and fast driving field for some high performance mechanical devices. Therefore, in applications where tunneling probability should be remaining small during the operation, the requirement on strong driving field can be relaxed by adding a few more resonators.

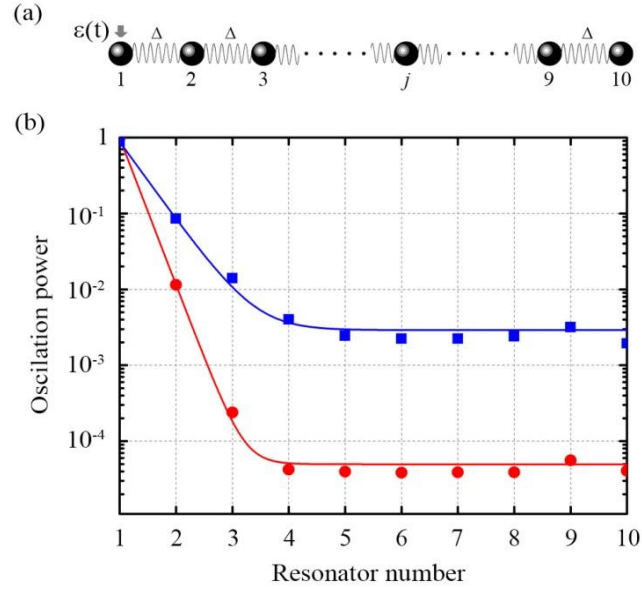

Figure 1S (a) scheme of mechanical lattice consisting of ten identical resonators. (b) Distribution of oscillation power at the DL condition. Oscillation power on each resonator is numerically calculated for the driving field with $A_d/2\pi = 375$ Hz, $\omega_d/2\pi = 166.2$ Hz (rectangle dots) and $A_d/2\pi = 1,000$ Hz, $\omega_d/2\pi = 419.3$ Hz (cycle dots) respectively. And the results are exponentially fitted (solid lines) to obtain the localization lengths.

## Supplementary References